\documentclass{aa}  

\usepackage{graphicx}
\usepackage{txfonts}

\begin{document} 

    \title{On The Detection of Digiorno-like Objects in the Flavor Zone}

   \author{
   Logan Pearce\inst{1}
          \and
          Sue D'oh Nym \inst{2}
          }

   \institute{Department of Astronomy, University of Michigan, Ann Arbor, MI 48109, USA\\
             \email{lapearce@umich.edu}
         \and
             Physics Department, Le Cordon Bleu, 888 Cheese and Sauce Rd, USA\\
             \email{whoami@youllneverknow.edu}
             }

   \date{Received April 1, 2026; accepted never}

 
  \abstract
   {}
   {This work proposes a new SETI search methodology under the assumption that a sufficiently advanced civilization could skip the middle man of converting starlight to energy to food preparation, and could directly harness their star's energy for food prep.}
   {We define the concept of the Flavor Zone (FZ): the optimal distance from a star for cooking food. To develop this definition we propose the toy model of a Digiorno-Like Object (DLO) and define the FZ as the regime for optimal cooking according to package directions.  We examine the effect of orbit on DLO cooking times and paradigms.  Finally, we study the feasibility of detection of DLOs in their FZs with current technology.}
   {We determined that DLOs aren't detecable with current technology nor should anyone ever try.}
   {}


   \maketitle
%

\section{Introduction}

The concept of the habitable zone (HZ) -- the region of separations from the host star for which liquid water may exist on the surface given sufficient atmospheric pressure -- in exoplanet science and astrobiology is central to the search for life on other worlds. The so-called ``Goldilocks Zone'', where it's not too hot and not too cold, defines a range of orbits for which the radiant energy received and resulting equilibrium temperature are within the range to support liquid water, thus making planets with orbits in that range amenable to hosting life as we know it, from microbial life to the possibility of advanced extraterrestrial civilizations.

Searches for evidence of advanced extraterrestrial civilizations, under the field of Search for Extraterrestrial Intelligence (SETI), often involve extrapolation of our own civilization (the only one we know of to date!) into observable signatures and our ability to detect them with current or near future technology.  More fanciful SETI ideas involve imagining new and creative ways a civilization may harness the power of their star to advance their civilization, such as the use of a Dyson sphere or ring to generate electricity to power their advanced devices and machines, including devices for food preparation. 

But we wondered, could a sufficiently advanced civilization skip the middle step of converting starlight into electricity into hot food, and directly harness their star's energy straight into food preparation?  The aim of this paper is to use a toy model, the Digiorno-like object (DLO), to investigate the feasibility of this method and the observational signals it would produce which might be detectable with current observational platforms. 


This paper is organized as follows.  In Section 2 we define our DLO toy model and the Flavor Zone (FZ).  In Sections 3 and 4 we investigate observational signatures of Flavor Zone transits and detectability with high-contrast imaging. In Section 5 we conclude that this was a huge waste of time.

\section{Definition of the Flavor Zone and Digiorno-like objects}

For our investigation we used the toy model of the Digiorno-like object (DLO). DLOs consist of a flat disk, diameters ranging from 10 to 16 inches, composed of tomato sauce and cheese on a bread dough, and typically some form of meat and veg toppings. They come in Rising Crust, Stuffed Crust, Croissant Crust, Hand-Tossed, Thin Crust, and Gluten-Free crust varieties\footnote{\url{https://www.goodnes.com/digiorno/products/}}. For our toy model we adopt a basic 12-inch (``medium") hand-tossed crust with cheese, pepperoni, sausage, green peppers, and onions (a typical ``supreme'' style)\footnote{While the authors did desire to adopt a ``Hawaiian'' pizza style model, we elected to avoid the controversy of pineapple pizza in favor of a more generally accepted pizza style. Even though Hawaiian pizzas are amazing.}\footnote{The authors are aware that olives typically come on a Supreme pizza, but the authors don't like olives.}

 \begin{figure}
   \centering
   \includegraphics[width=0.49\textwidth]{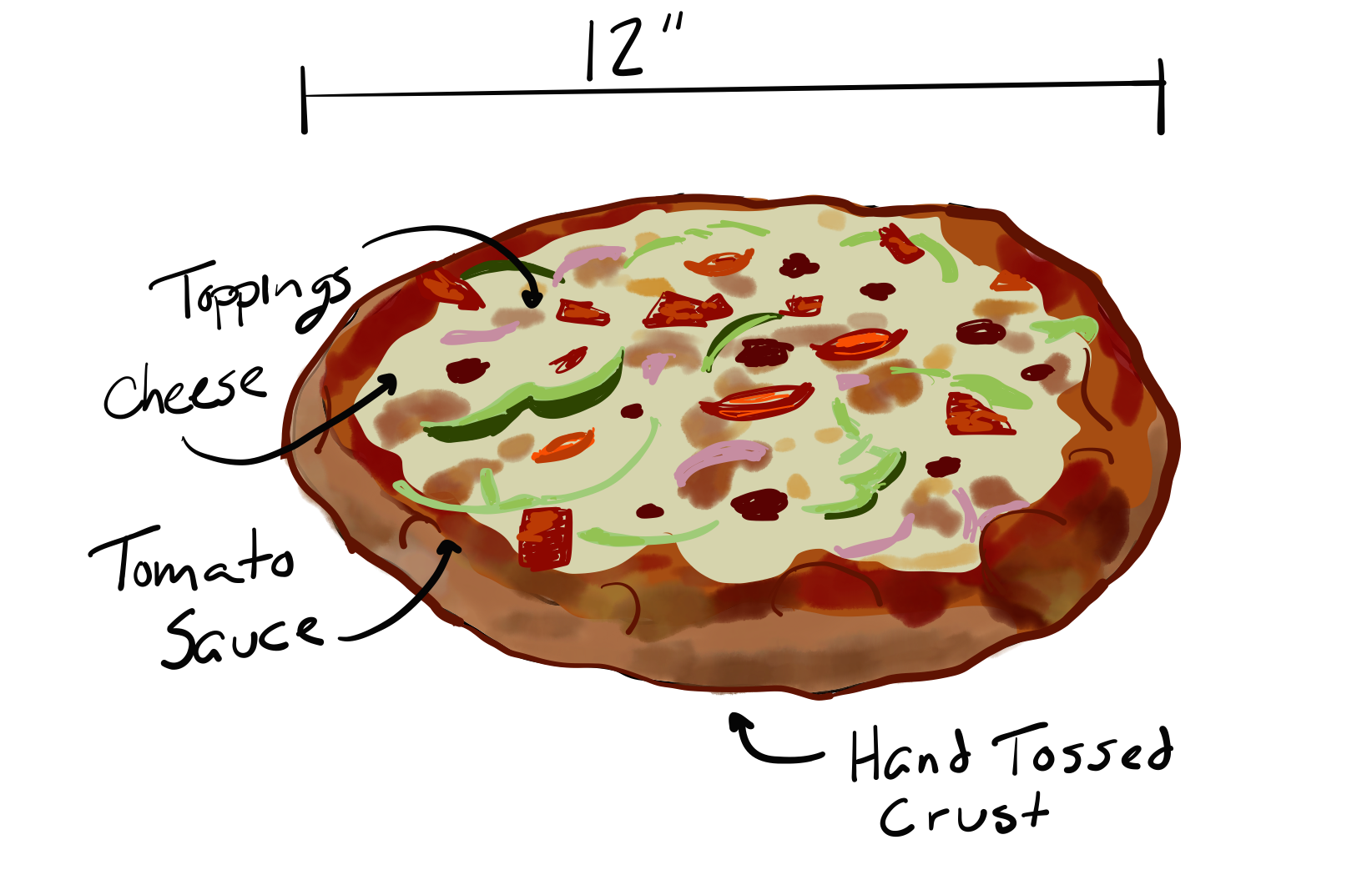}
   \caption{The structure of our model of a Digiorno-like Object.}
              \label{}%
    \end{figure}

To derive the definition of the Flavor Zone (FZ) we adopt the typical definition of equilibrium temperature:
\begin{equation}
    P_{\rm{incident,\odot}} = L_\odot \left( \frac{\pi R^2_{\rm{DLO}}}{4 \pi r^2} \right)
\end{equation}
where $P_{\rm{incident,\odot}}$ is the incident power on the DLO by the host star, $L_\odot$ is the luminosity of the star, where $L_\odot = 4 \pi \sigma_{\rm{SB}} R_\odot^2 T_\odot^4$, $R_{\rm{DLO}}$ is the DLO radius, and $r$ is the star-DLO separation.  Then the energy absorbed by the DLO is:
\begin{equation}
    P_{\rm{absorbed,DLO}} = P_{\rm{incident,\odot}} \times (1 - A)
\end{equation}
where $A$ is the albedo. Then, the power emitted by the DLO is:
\begin{equation}
    P_{\rm{emitted,DLO}} = 4 \pi \sigma_{\rm{SB}} R^2_{\rm{DLO}} T^4_{\rm{DLO}}
\end{equation}
In equilibrium $P_{\rm{emitted}} = P_{\rm{absorbed,DLO}}$:
\begin{equation}
    T_{\rm{DLO}} = \left[ \frac{(1-A)}{4\;r^2} R_\odot^2 T_\odot^4 \right]^{\frac{1}{4}}
\end{equation}

The Digiorno website recommends baking your DLO at 400$^\circ$~F (477.6~K) for 22-25 min. Solving for separation:
\begin{equation}
    r = \left[ \frac{(1 - A) \;R_\odot^2 \; T_\odot^4}{2.1\times10^{11}\;K} \right]^{\frac{1}{2}}
\end{equation}
Given the difficulties and intricacies involved in maintaining a precise distance for 20-25 minutes, we adopt a narrow temperature range of 375--425$^\circ$~F (463.7--491.5~K) as acceptable for the inner and outer boundaries of the Flavor Zone.

To estimate the albedo of a DLO we used Figure 1 of Frank \& Birth 1982, which shows the reflectance of cheddar and parmesan cheeses as a function of wavelength. The albedo ranges from $\sim$0.6 at 0.95~$\mu$m to $\sim$0.03 at 1.45~$\mu$m. We adopted the cheddar curve from that figure, and even though the albedo seems to be very wavelength dependent, for our toy model we adopted a uniform A~=~0.06 as the approximate albedo value for our DLOs.

Table \ref{table:FZ} shows the FZ for several nearby stars including the Sun. The nearby brown dwarf has the nearest FZ, while the hot star Sirius A has the widest FZ in the neighborhood.

\begin{table*}
\caption{Flavor Zone locations for several nearby stars} 
\label{table:FZ}      
\centering                        
\begin{tabular}{c c c c}  
\hline\hline           
\vspace{3mm}
Star & SpT, T$_{\rm{eff,\odot}}$ [K], R$*,_\odot$ & Flavor Zone [au] & Flavor Zone [mas] \\   
\hline                    
   The Sun & G2V, 5780, 1 & 0.31 -- 0.35 & NA \\      
   Proxima Centauri & M5.5V, 2992, 0.15 & 0.013 -- 0.014 & 9.7 -- 10.9 \\
   Alpha Centauri A & G2V, 5790, 1.2 & 0.37 -- 0.42 & 289.4 -- 325.1 \\
   Wolf 359 & M6.0V, 2794, 0.14 & 0.010 -- 0.011  & 4.25 -- 4.78 \\
   Sirius A & A1V, 9940, 1.7 & 1.6 -- 1.8    & 594.9 -- 668.4 \\
   van Maanen's Star & DZ7, 6130, 0.011 & 0.0038 -- 0.0043   & 0.89 -- 1.00 \\ 
   Boyajian's Star & F3V, 6750, 1.58 & 0.67 -- 0.76 & 1.5 -- 1.7 \\
   Luhman 16 A & L8, 1350, 0.087 & 0.0015 -- 0.0017 & 0.74 -- 0.83 \\
\hline                                  
\end{tabular}
\end{table*}

\subsection{The problem of DLO orientation}

\begin{figure}
   \centering
   \includegraphics[width=0.4\textwidth]{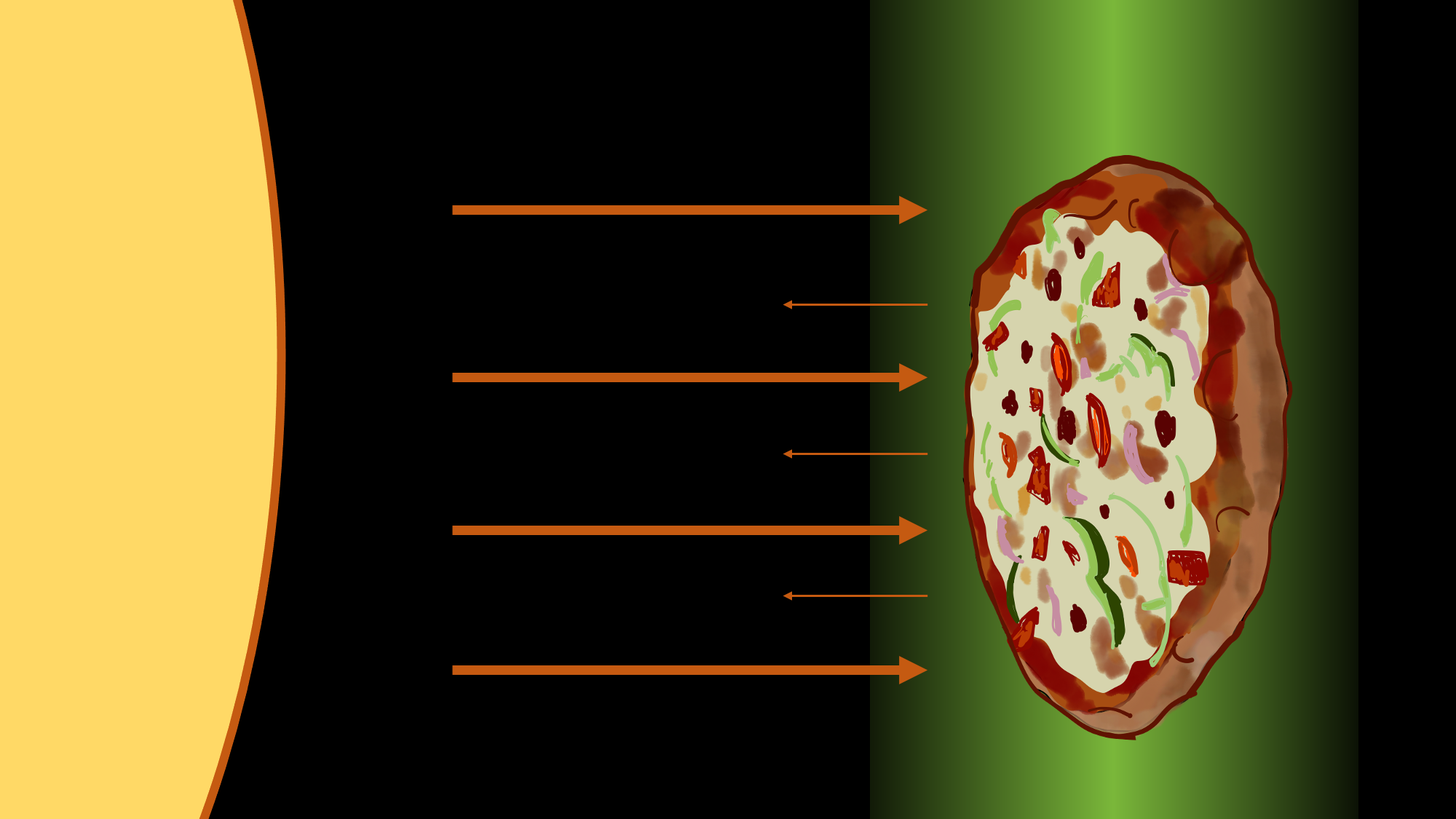}
   \includegraphics[width=0.4\textwidth]{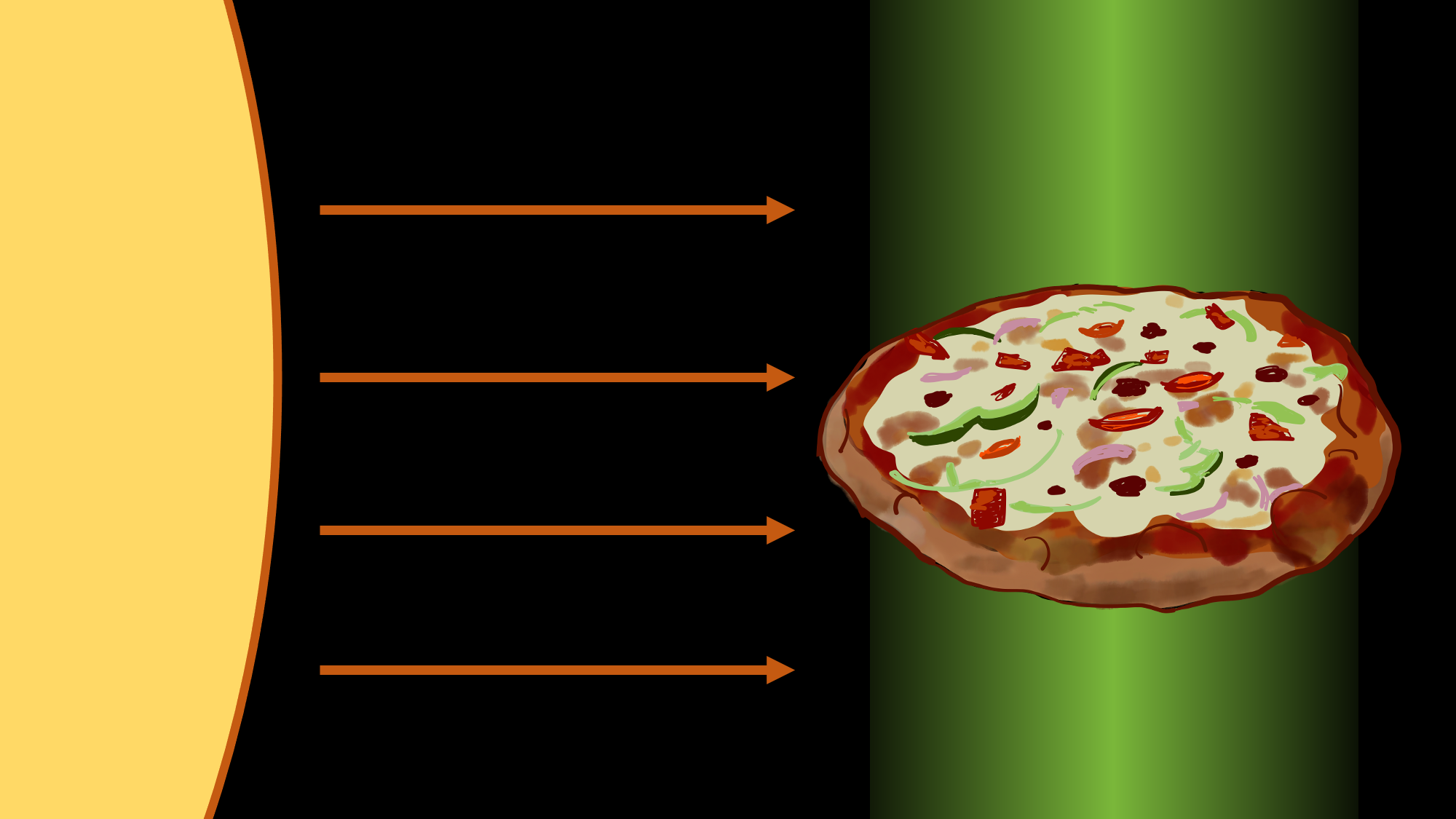}
   \caption{DLO orientation. Top: cheese-on orientation.  Bottom: crust-on orientation}
              \label{fig:DLO-orient}%
    \end{figure}

Unlike an oven, the orientation of the DLO in relation to the heat source must be considered since energy is coming from one direction only. Figure \ref{fig:DLO-orient} shows the two extremes of orientation options, one for which the DLO is oriented face on (cheese-on) to the star and one which is edge on (crust-on).  The FZ is marked with green but is not to scale.  

Orienting the DLO cheese-on is clearly ideal as it maximally exposes the tasty bits to the heat source.  However, only the tasty bits side is facing the heat while the crust remains exposed to the cold of space, thus we estimate that cooking times will need to be expanded with a 180$^{\circ}$ flip needed approximately halfway through.

Orienting the DLO crust-on presents a few challenges.  One, the area exposed directly to the heat source is much reduced.  In the cheese-on orientation an area of $A = \pi\;r^2 = 113$in$^2$ is exposed to the sunlight, while in the crust-on orientation and area of $A = \rm{width} \times \rm{height} = 12 \times 1 = 12$in$^2$ is exposed, greatly reducing the energy able to be absorbed. Power $\propto$ Area, so in the crust-on orientation the DLO receives 10\% of the ergs per second of cheese-on. Over the 25-minute cooking time the crust-on DLO will have received 160 fewer ergs of energy than cheese-on, greatly extending the needed cooking time.  Also multiple flips and spins would likely be required for even cooking.  

An additional concern we considered was the difference in temperature between the close edge and the far edge in the crust-on orientation, however this turns out to be negligible. In the case of Proxima Centauri's FZ the far edge would receive $3.1\times10^{-10} \rm{ergs}\;s^{-1}$ less than the leading edge, given that Power~$\propto \frac{1}{r^2}$ and the 12" diameter of the model DLO.  Over 25 minutes cooking time the far edge would receive $4.6\times10^{-7} \rm{ergs}$ fewer than the close edge, hardly something worth considering here.

In summary, we assumed that the cheese-on orientation is optimal for maximal cooking efficiency and used that orientation for our model. We also considered that the alien chefs might want to attach a thruster pack to their DLO to aid with orientation maintenance, but neglect the thruster pack in our model.

\subsection{The orbit}

While cooking, the DLO will need to maintain an orbit around its host star for at least the cooking time if not longer.  This means that it must be on an orbit such that it remains in the FZ throughout the cooking time.  The most obvious solution is to place the DLO squarely within the FZ on a perfectly circular orbit (eccentricity = 0.0).  This requires initialzing the DLO on the orbit with the precise velocity for a circular orbit:
\begin{equation}
    v_{\rm{circ}} = \sqrt{\frac{G\; m_{\rm{tot}}}{a}}
\end{equation}
where $m_{\rm{tot}}$ is the total mass of the star--DLO system (which in this case is the mass of the star since the DLO mass is negligible), $G$ is the gravitational constant, and $a$ is the semi-major axes of the orbit.  For Proxima Centauri, the center of the FZ is at 0.013~au, the mass is 0.1221~M$_\odot$, and the circular velocity is 90.25 km s$^{-1}$.

However intializing a DLO on the precise correct circular velocity to maintain a circular orbit is difficult. There are any number of ways the velocity may be too high or too low initially, or change over time. So we examined the range of orbital parameters acceptable to produce a perfectly cooked DLO.

\begin{figure}
   \centering
   \includegraphics[width=0.4\textwidth]{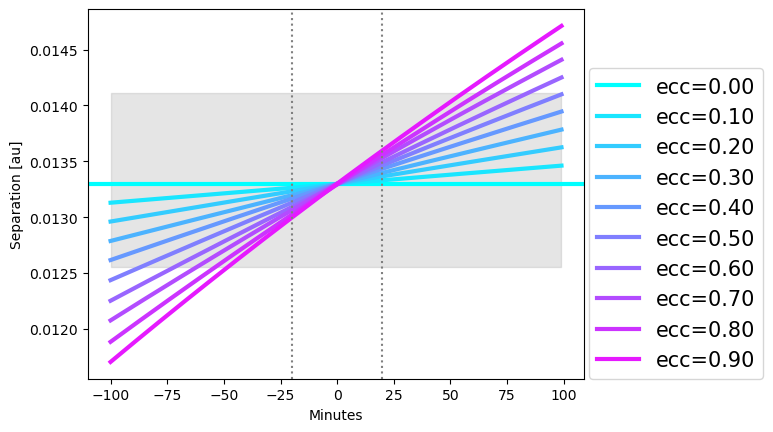}
   \includegraphics[width=0.4\textwidth]{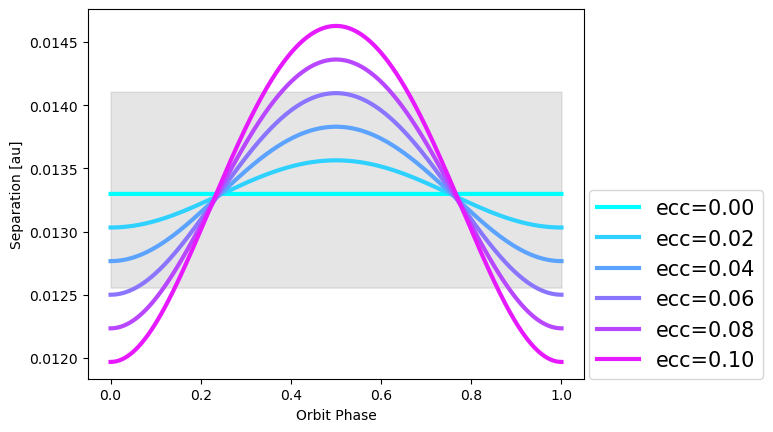}
   \caption{Orbits for a DLO around Proxima Centauri for varying eccentricities. Semi-major axis is set to the middle of the FZ; grey shaded region shows the FZ region, grey dashed lines show release +/- 20 minutes cooking time.  Top: Separation as a function of time for eccentricities 0--1.0. Bottom: Entire orbit for eccentricities 0--0.10. Takeaway: All orbits remain within the FZ for an entire 20 minute cooking timeif released in the center of the FZ, but only orbits with ecc $<$ 0.06 remain in the FZ for an entire orbit. }
              \label{fig:orbits}%
    \end{figure}

Figure \ref{fig:orbits} shows orbits of different eccentricities and the star-DLO separation over time. The top plot shows orbits of varying eccentricities over time for Proxima Centauri's FZ, assuming the DLO is released at the exact center of the FZ (0.0132 au). The grey shaded region shows the inner and outer FZ boundaries as stated above, using 375-425$^\circ$F as the boundaries; the dashed grey lines show +/- 20 minutes from the release. All orbits remain within the FZ during the 20 minute intervals on either side of release. The bottom plot shows the entire orbit for eccentricities from 0.0 -- 0.10; only orbits with eccentricity $<$0.06 remain within the FZ during the entire orbit. Given the above discussion of the need to extend cook times with one or more flips during cooking, and the likelihood that alien chefs may want to cook more than one DLO at a time, staying on an orbit that remains entirely within the FZ would be recommended.

\subsection{The problem of solar radiation pressure}

Given the small mass of the DLO, solar radiation pressure (SRP) will likely be a non-negligble external perturbation to the orbital elements, even during the relatively short cooking times, and could possibly push the DLO outside the FZ.  Someone could definitely do the math on that, but we've already done a lot of analysis of the orbit for this ridiculous paper, don't you think?  If you care, the effect of SRP on the DLO orbit is left as an exercise to the reader.

\section{Simulations of Flavor Zone transits of Digiorno-like Objects}

So how could an enterprising SETI scientist detect the presence of a DLO?  How about in a transit light curve?

\begin{figure}
   \centering
   \includegraphics[width=0.4\textwidth]{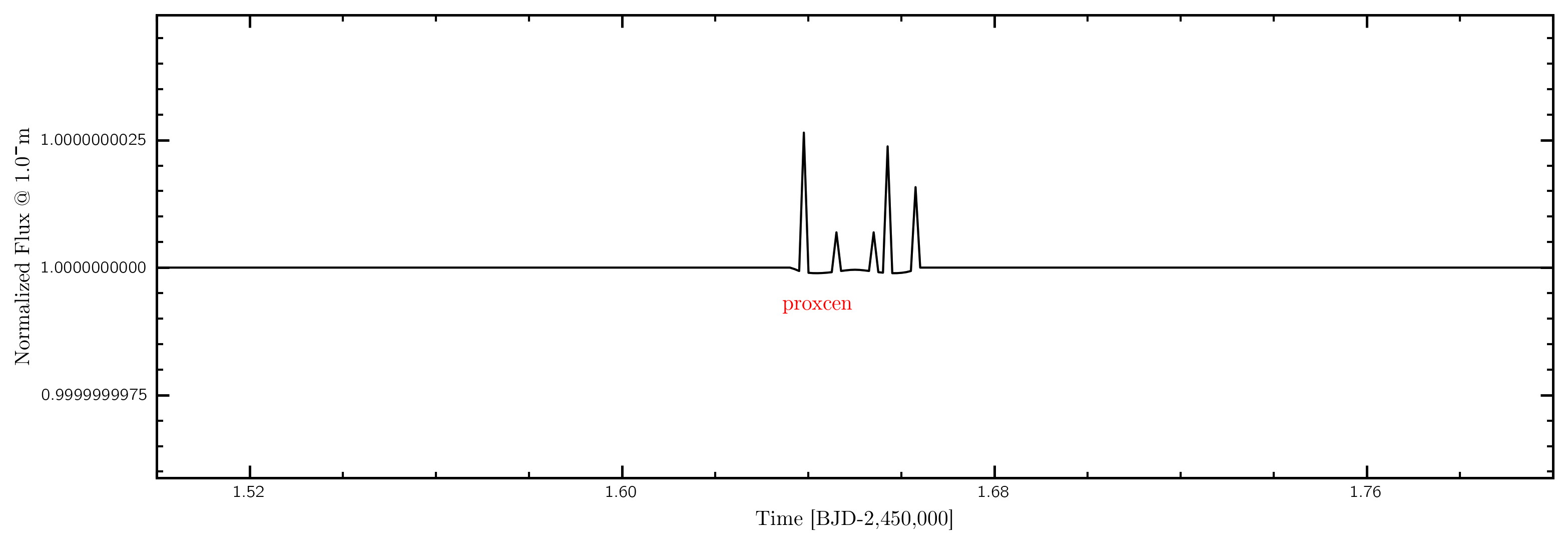}
   \includegraphics[width=0.4\textwidth]{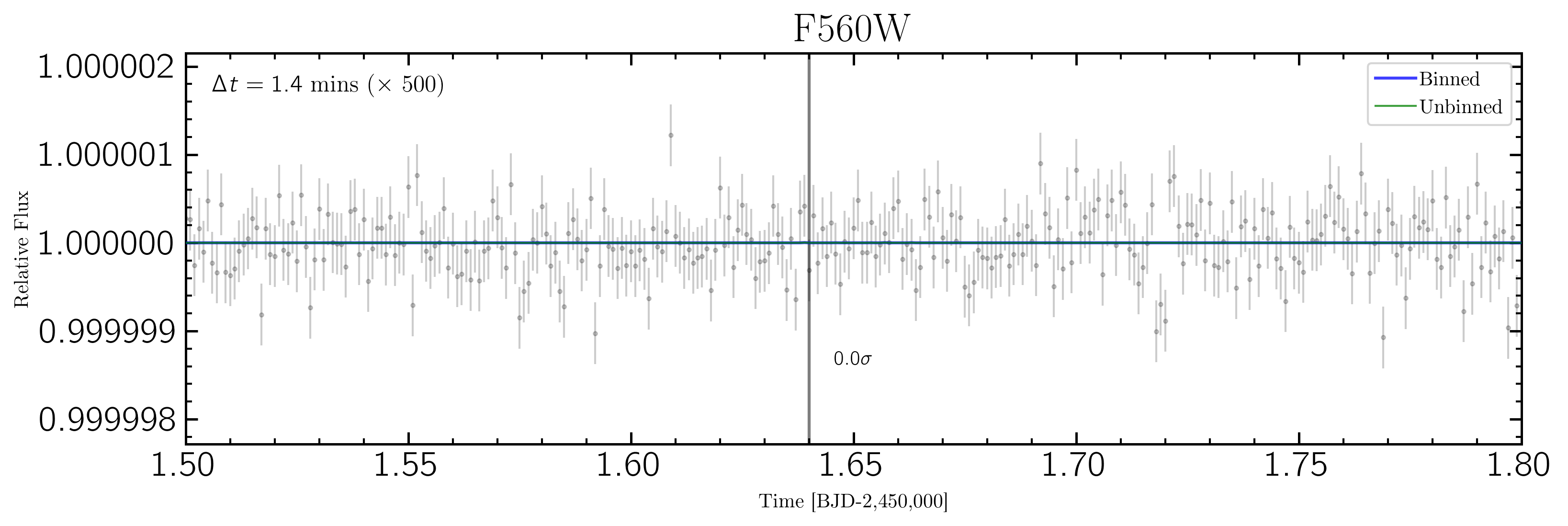}
   \caption{Transit simulations of our DLO model around Proxima Centauri. Top: \texttt{planetplanet} transit simulation.  Bottom: Simulated JWST observation of DLO trasnit}
              \label{fig:transit}%
    \end{figure}

We used the python package \texttt{planetplanet} (Luger et al. 2017) to model a DLO transiting the nearest star Proxima Centauri.  Using Proxima Centauri's mass, radius, and distance, along with a DLO of radius 6 inches (2.4e-8 R$_{\odot}$) and mass 900 grams (1.5e-25 M$_{\odot}$).  Figure \ref{fig:transit} (top) shows the resulting transit light curve.  The dip in flux is as small or smaller than the resolution of the simulation.  Figure \ref{fig:transit} (bottom) shows a simulated JWST observation in the F560W filter, with no transit visible.

It seems that trying to detect a DLO via transit light curves is, unsurprisingly, not possible with current technology. But once we can detect a flux dip of 1 part in 1 billion, that will be golden DLO searching time.

\section{High-Contrast Imaging Requirements of Flavor Zone Objects}

Perhaps within the high-contrast imaging (HCI) field we will find some hope.

Examining Table \ref{table:FZ} we see that for cool stars the separations required for the FZ are of the order 10 mas or smaller, putting them way out of reach of resolving with HCI, however separations of Alpha Cen A and Sirius A are favorable to HCI techniques.

\begin{figure}
   \centering
   \includegraphics[width=0.49\textwidth]{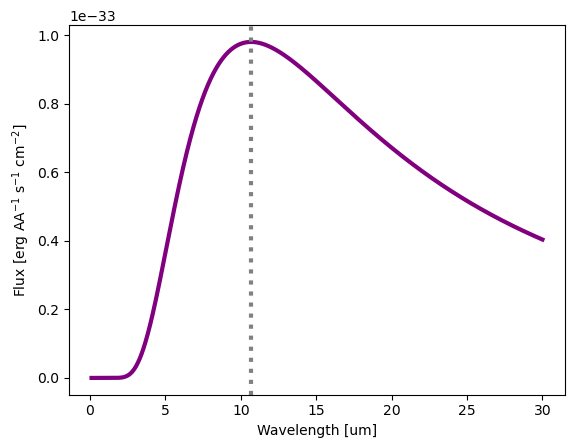}
   \caption{Blackbody spectrum for an object at $T~=~477.6$~K at the distance of Proxima Centauri with the size of our DLO model.}
              \label{fig:bb}%
    \end{figure}

\textit{Thermal Emission?} The thermal flux emitted by a fully cooked DLO we assumed to be a blackbody with $T~=~477.6$~K. We computed the blackbody flux for an object the size of our DLO model at a distance of Alpha Centauri. The peak of the blackbody occurs at $\lambda = 10.7\mu m$, shown in Figure \ref{fig:bb}. At $10.7\mu m$, the DLO blackbody flux is 10e-34 Ergs cm$^{-2}$ s$^{-1}$ Ang$^{-1}$, while Alpha Centauri's flux is 7.9e-11 Ergs cm$^{-2}$ s$^{-1}$ Ang$^{-1}$, giving a contrast ratio of 1e-23 or 57 magnitudes. This is still beyond the reach of even the best IR imagers.

\textit{What about reflected light?} In Frank and Birth 1981, Fig 1, the best case cheese albedo is about A = 0.6 for optical wavelengths, which is optimal for reflected light imaging, so we adopted this albedo and assumed no wavelength dependence for simplicity.

\begin{figure}
   \centering
   \includegraphics[width=0.49\textwidth]{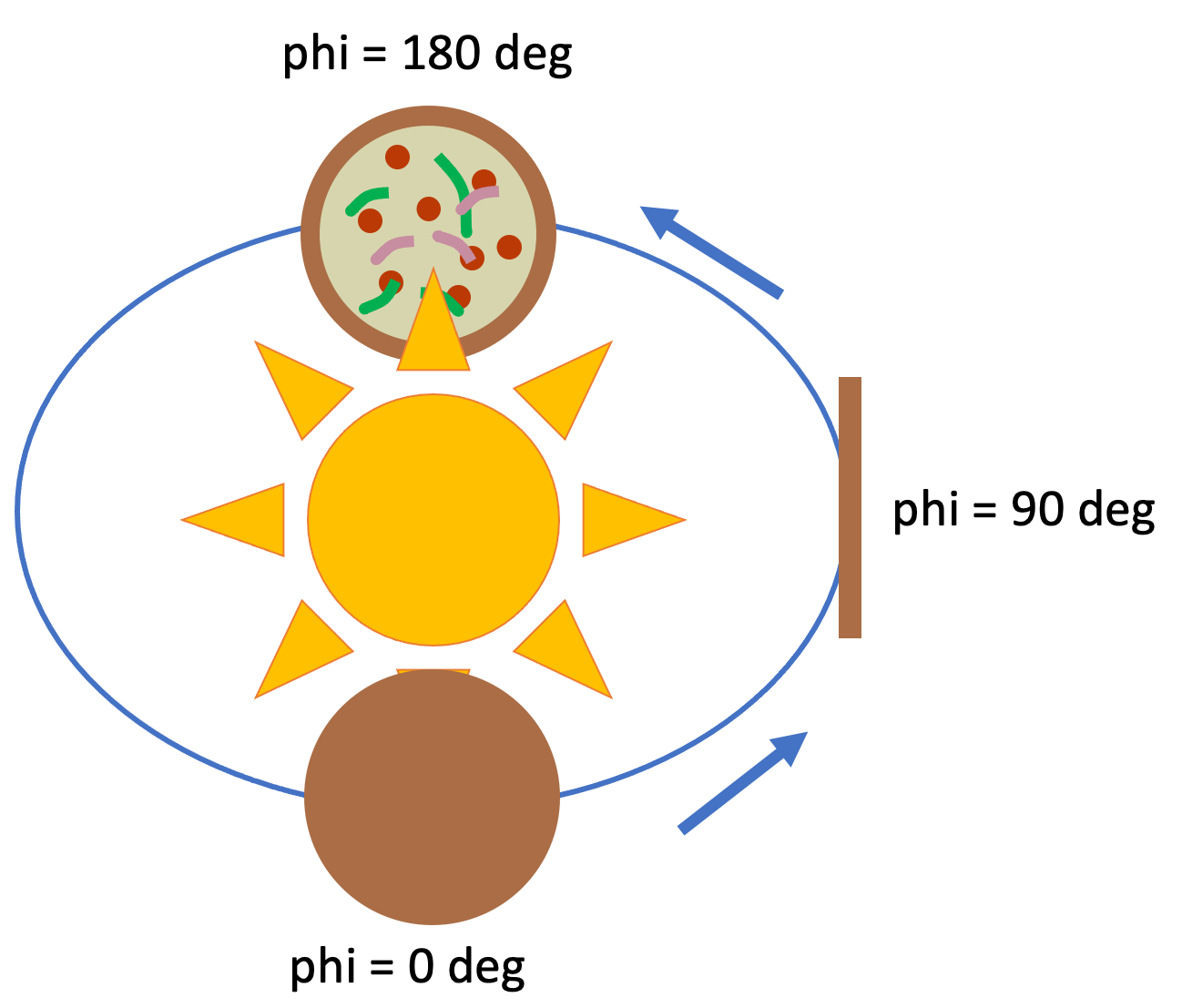}
   \caption{Illustration of reflected light phase of a DLO as it goes through an orbit orbit. Phi is the orbit phase. Not to scale.}
              \label{fig:phase}%
    \end{figure}

Figure \ref{fig:phase} illustrates the orbit phase, $\phi$, of a DLO through an orbit. Orbit phase $\phi = 0^\circ$ corresponds to the inferior conjunction (DLO aligned between star and observer) and increasing through orbit to max phase at superior conjunction (star between observer and DLO) at $\phi = 180^\circ$.  Assuming a Lambertian phase function for a uniform refelcting sphere, the observed phase $\alpha$ depends on inclination, and is given by $\cos(\alpha) = \sin(i)\cos(\phi)$, where $i$ is orbital inclination.  Reflected light flux contrast for a sphere then is given as a function of phase by: $C(\alpha) = A_g \left(\frac{R_p}{\rho}\right)^2 \frac{1}{\pi} \left[\sin(\alpha) + (\pi - \alpha)\cos(\alpha) \right]$.  Thus the brightest an object will be will be at superior conjunction (full phase for an edge-on orbit, where the object is directly behind the star), half illuminated at quadrature, and at its darkest at inferior conjunction.  For a face-on orbit the object will be at quadrature for the entire orbit.  

But, a sharp reader will have noted, a DLO is not a sphere.  The DLO will still be brightest at superior conjunction and darkest at inferior conjunction, where the DLO most resembles a circle.  However, for nearly edge-on orbits, the DLO is also very close to or behind the star at superior conjunction.  Quadrature is typically the best time to observe a sphere in reflected light for this reason, however the DLO will appear as only a small sliver on the sky to an observer at this angle.  Essentially no light will be reflected to the observer when the DLO is at quadrature.  So our only hope is for any phases $>~90^\circ$.

But the contrast also depends on the radius and separation of the reflecting object.  For the Alpha Centauri case, and our 12-inch DLO model, at full phase the reflected light contrast is $C = 3.9\times 10^{-24}$, or almost 60 magnitudes.  

Wow.  That's even worse than the thermal case.

\section{Conclusion}
In this work we have examined the SETI detection potential of a civilization using their star's energy to directly heat and prepare food.  We examined the toy model of a Digiorno-like object and its detectability with current technology.  We defined the Flavor Zone, the region of optimal temperatures for fully cooking a DLO.  We determined that DLOs in the FZ of nearby stars are well below detection limits for current technology.  Transit signals of 1 part in a billion are needed for DLO transit detections, and contrasts of order 60 magnitudes in optical or IR wavelengths are needed for direct detection.

\subsection{Future Work and Recommendations}
Recommendations: Do not attempt to search for ETI in the form of DLOs.

\noindent Future work: None. Don't do this.

\begin{acknowledgements}
       L.A.P acknowledges nobody because they did this all on their own for better or worse.  S.D.N acknowledges no one because I mean really who would want to be associated with this mess?  \\
      
      No one funded this research nor should they.
\end{acknowledgements}

\noindent \textbf{REFERENCES}\\
We do not include formal references under the assumption that authors of those real, legitimate works don't want their work tied to this nonsense.
%
%

\end{document}